\documentclass[preprint,prd,superscriptaddress,nofootinbib,floatfix]{revtex4}

\usepackage{amsmath,amssymb}
\usepackage{graphicx}
\usepackage{wrapfig}
\usepackage{bm}
\usepackage{multirow}
\usepackage{color}

\begin{document}

\preprint{OU-HET-816}
\preprint{KEK-CP-306}
\preprint{YITP-14-38}

\title{
  Computation of the electromagnetic pion form factor from lattice QCD in the $\epsilon$ regime
}

\newcommand{\Tsukuba}{
  Graduate School of Pure and Applied Sciences,
University of Tsukuba, Tsukuba, Ibaraki 305-8571, Japan
}

\newcommand{\CCS}{
Center for Computational Sciences,
University of Tsukuba, Tsukuba, Ibaraki 305-8577, Japan
}
\newcommand{\KEK}{
  KEK Theory Center,
  High Energy Accelerator Research Organization (KEK),
  Tsukuba 305-0801, Japan
}
\newcommand{\GUAS}{
  School of High Energy Accelerator Science,
  The Graduate University for Advanced Studies (Sokendai),
  Tsukuba 305-0801, Japan
}
\newcommand{\YITP}{
  Yukawa Institute for Theoretical Physics, 
  Kyoto University, Kyoto 606-8502, Japan
}
\newcommand{\Osaka}{
  Department of Physics, Osaka University,
  Toyonaka 560-0043, Japan
}

\author{H.~Fukaya}
\affiliation{\Osaka}

\author{S.~Aoki}
\affiliation{\YITP}
\affiliation{\CCS}


\author{S.~Hashimoto}
\affiliation{\KEK}
\affiliation{\GUAS}

\author{T.~Kaneko}
\affiliation{\KEK}
\affiliation{\GUAS}

\author{H.~Matsufuru}
\affiliation{\KEK}
\affiliation{\GUAS}

\author{J.~Noaki}
\affiliation{\KEK}



\collaboration{JLQCD collaboration}
\noaffiliation


\begin{abstract} 
We calculate the electromagnetic pion form factor
in lattice QCD with $2+1$ flavors of the dynamical overlap quarks.
Up and down quark masses are set below their physical values 
so that the system is in the so-called $\epsilon$ regime
with the small size of our lattice $\sim 1.8$ fm.
The finite volume corrections are generally expected to be $\sim 100 \%$
in the $\epsilon$ regime. 
We, however, find a way to automatically cancel the dominant part of them.
Inserting non-zero momenta
and taking appropriate ratios of the two and three point functions, 
we can eliminate the contribution from the zero-momentum pion mode.
Then the remaining finite volume effect is a small perturbation
from the non-zero modes.
Our lattice data agree with this theoretical prediction and
the extracted pion charge radius is consistent with the experiment.
\end{abstract}
\maketitle

\section{Introduction}
\label{sec:intro}
Dynamics of pions is governed by its nature as the Nambu-Goldstone boson
associated with the spontaneous breaking of chiral symmetry in the
vacuum of Quantum Chromodynamics (QCD).
Beyond the leading order in the expansion in terms of pion momentum
squared $p^2$ and pion mass squared $m_\pi^2$, it develops
non-analytic functional form \cite{Gasser:1983yg, Gasser:1984gg},
known as the chiral logarithm.
For the charged pion form factor $F_V(q^2)$ as a
function of the momentum transfer $q$, in particular, 
the charge radius defined by 
\begin{equation}
  \label{eq:rV2}
  \langle r^2\rangle_V\equiv 6 
  \left.\frac{\partial F_V(q^2)}{\partial q^2}\right|_{q^2=0}
\end{equation}
is predicted to diverge in the limit of vanishing pion mass,
{\it i.e.} $\sim\ln m_\pi^2$.
In order for numerical computations of lattice QCD to be reliable in
reproducing the low-energy property of the pions, it is crucial to
confirm this remarkable behavior.

In the lattice QCD simulations, approaching the chiral limit is
challenging because the computational cost to invert the Dirac
operator grows as $1/m_\pi^2$.
Furthermore, finite volume effect is expected to increase as $m_\pi$ decreases.
Therefore, to avoid large systematic effect from the volume,
the overall cost increases much faster than $1/m_\pi^2$.
Most of the previous calculations, including our own work \cite{Aoki:2009qn},
have been performed at large pion masses ($\gtrsim$ 300~MeV), 
and the results for the pion charge radius were significantly lower than the experimental value.
Recent works \cite{Nguyen:2011ek, Brandt:2013dua, Koponen:2013boa} are simulating lighter pions and the results are
indeed showing an increase towards the physical pion mass. 
However, in the vicinity of the chiral limit, the violation of chiral symmetry
becomes an issue with the conventional lattice fermion formulations,
such as the Wilson fermions.
They violate the chiral symmetry at the order of
$a^2 \Lambda_{\rm QCD}^3$
(assuming the $O(a)$-improved action),
where $\Lambda_{\rm QCD}$ ($\sim$ 300~MeV)
is the typical scale of QCD.
In the most recent dynamical simulations, the lattice cutoff $1/a$ is
around 3~GeV, and the size of the violation is thus an order of 3~MeV,
which is only slightly below the physical up and down quark masses.
This implies that in the quark mass regime we are interested in, 
the violation of chiral symmetry due to the lattice artifact is 
as large as in magnitude the effect due to the quark mass.
Discretization effect in such a situation could become sizable.

In this work we carry out a lattice calculation of the pion charge
radius near the chiral limit using the fermion formulation that
preserves exact chiral symmetry
\footnote{
Our preliminary results were presented in Ref.~\cite{Fukaya:2012dla}.
}.
We simulate lattice QCD with up and down quark masses below 
the physical point, employing the overlap fermion action
\cite{Neuberger:1997fp,Neuberger:1998wv}.
This overlap fermion action has an exact chiral symmetry
through the Ginsparg-Wilson relation \cite{Ginsparg:1981bj, Luscher:1998pqa}. 
In our numerical implementation, this relation is kept
at the level of $10^{-8}$ accuracy.
Therefore, the chiral logarithm is expected to have
the same functional form as the continuum theory.
By obtaining a data point with an extremely small pion and combining
it with our previous results at a larger mass region,
we make an interpolation into the physical point.
This {\it chiral interpolation} can suppress
the systematic error from the mass dependence of the data, 
and confirm if there is the expected divergent behavior of the pion charge radius.

At the small quark mass ($\sim$ 3~MeV), the finite volume effect is
expected to be quite large for the lattice size ($\sim$ 1.8~fm) that we use.
It is often mentioned in the literature \cite{Aoki:2013ldr} that
$m_\pi L$ has to be larger than 4 in order to suppress the finite
volume effect at a few per cent level or lower.
At our simulated pion mass, $L$ must be as large as 5--6~fm 
to satisfy this criterion.
In the regime where the near-zero modes determine the dynamics,
however, the dominant part of  the finite volume effects come from 
the zero-momentum mode of the pions, while 
the higher energy states give still exponentially small effects
(Note that the energy of non-zero momentum modes in a finite volume 
satisfies $E_\pi> 2\pi/L$, and thus, $E_\pi L>6$ ).
Therefore, once we remove the effect of the pion zero-momentum mode, 
the remaining finite volume effect is manageable.
For this purpose, the so-called $\epsilon$ expansion 
\cite{Gasser:1986vb, Gasser:1987zq, Neuberger:1987zz, Leutwyler:1992yt}
was developed and
applied to extract the low-energy constants of chiral perturbation theory (ChPT).

The $\epsilon$ expansion is valid for a system
where the pion Compton wavelength exceeds $L$.
In this $\epsilon$ regime, the zero momentum mode may rotate in the
flavor group manifold $SU(N_f)$ 
and therefore, should be treated non-perturbatively.
Such analysis leads to the prediction of 
the low-lying Dirac operator eigenvalue spectrum 
\cite{Shuryak:1992pi, Verbaarschot:1993pm, Verbaarschot:1994qf, Damgaard:1998xy, Damgaard:2008zs},
as well as the pseudoscalar two-point functions \cite{Hansen:1990un, Aoki:2011pza}.
These formulas have rather complicated expressions containing Bessel functions,
but nicely describe the lattice data 
\cite{DeGrand:2006nv, Hasenfratz:2007yj, Lang:2006ab, Fukaya:2007fb,Fukaya:2007yv,Fukaya:2007pn,
Hasenfratz:2008ce,Fukaya:2009fh,
Fukaya:2010na,Bar:2010zj, Bernardoni:2010nf},
and are useful to determine the leading two low-energy constants,
the chiral condensate $\Sigma$, and pion decay constant $F$.

In order to extract the pion form factor from the $\epsilon$ regime
lattice calculation, we need the ChPT prediction
of the three-point functions, which is not known in the literature,
except for the kaon sectors \cite{Hernandez:2002ds, Giusti:2004an, Hernandez:2008ft}.
Even if such predictions were available, 
the analysis would require a non-trivial task to 
disentangle the low-energy physics from some complicated
form of the Bessel functions coming from the zero-mode.

In this work, we would like to show a new direction,
using the $\epsilon$ expansion in a more indirect way.
Namely, we use the $\epsilon$ expansion of ChPT just for
finding the combination of the correlators which has a small sensitivity to the volume.
We find that this is possible by inserting non-zero momenta 
to the relevant operators (or simply taking differences of them at different time-slices),
and taking appropriate ratios of them.
This procedure automatically eliminates the leading ${\cal O}(1)$
finite volume effects, and 
the remaining next-to-leading order contributions are
expected to be a small perturbation \cite{Fukaya:2014uba, Suzuki}.

This method considerably simplifies the analysis in the $\epsilon$ regime.
Since the dominance of the pion zero-mode contribution in the finite
volume effect is universal for most correlators, we expect that the 
application of the method is wider, {\it e.g.}
other meson/baryon form factors.
Even in the $p$ regime, our method suggests a way to minimize the
finite volume effects.
In this work, we present the result for the electro-magnetic pion form
factor as the first example.
Our lattice data for the electro-magnetic form factor agree with this
theoretical expectation reasonably well, yielding a consistent value
of the pion charge radius with the experiment. 

This paper is organized as follows.
In Sec~\ref{sec:2pt}, we revisit the 
two-point functions in the $\epsilon$ regime
of ChPT and demonstrate how our new strategy
works to automatically cancel the dominant
finite volume effects of the pion zero-mode.
In Sec.~\ref{sec:3pt}, we compute the $\epsilon$
expansion of the three-point functions 
and find the ratios of the correlators
which are free from the pion zero-mode's contamination.
The result of our simulation is presented
in Sec.~\ref{sec:lattice} and we give
our conclusion in Sec.~\ref{sec:conclusion}.

\section{Two-point functions in the $\epsilon$ regime}
\label{sec:2pt}

Let us consider the two-point correlators to illustrate our idea.
For simplicity, we consider two-flavor ChPT with a degenerate
quark mass $m$ in a finite volume $V=L^3 T$, 
of which boundary condition is set periodic in every direction.
Let us denote the chiral condensate by $\Sigma$ 
and the pion decay constant by $F$.
Including the (sea) strange quark is not difficult
\cite{Bernardoni:2008ei} and does not change the following results 
at the leading-order of ChPT.

In the $\epsilon$ expansion of ChPT,
the pion's zero-momentum mode is exactly treated by performing a group
integral over $SU(N_f)$, where $N_f=2$ is the number of flavors,
while the non-zero modes and their interactions are
perturbatively treated. 
Namely, it is a (weakly coupled) system of $SU(2)$ matrix model 
(or a $U(2)$ matrix model when the global topological charge of the
gauge field is fixed) and massless fields.

A two-point correlation function of pseudo-scalar density operator
$P(x)$ separated by a four vector $x=(t, x_1,x_2,x_3)$ 
is expressed as \cite{Bernardoni:2008ei}
\begin{eqnarray}
  \label{eq:PP}
\langle P(x)P(0) \rangle &=& X 
+ Y \left(\frac{1}{V}\sum_{p \neq 0} \frac{e^{ipx}}{p^2}\right)
+Z \left(\frac{1}{V}\sum_{p \neq 0} \frac{e^{ipx}}{(p^2)^2}\right)+\cdots ,
\end{eqnarray}
where $X$, $Y$, $Z$, ... are non-trivial (Bessel) functions of 
$m \Sigma V$ arising from the zero-mode integrals.
Unlike the conventional meson propagator, there is a constant term $X$,
which is a contribution purely from the zero-mode.
The second and third terms represent the coupled contribution of the
zero modes to the non-zero modes described as a massless scalar field.
It is massless because the mass term is a small perturbation in the
$\epsilon$ expansion. 
Note that the $p=0$ part contribution is
absent in the momentum summations and 
this expression is manifestly free from infra-red divergences.

Since the non-zero modes are treated as massless bosons,
the correlation function projected onto zero spatial momentum becomes 
a polynomial function of $t$, which is a remarkable difference from 
the conventional exponential function $\exp(-m_\pi t)$ 
in the conventional $p$ regime.
In fact, this special property of the $\epsilon$ expansion,
was used to extract the low-energy constants from finite volume
lattice QCD
\cite{Fukaya:2007fb,Fukaya:2007pn,Fukaya:2009fh,Fukaya:2011in}.  
In this work we try to avoid the terms arising from the zero-mode
integral, which is characteristic of the $\epsilon$ regime. 

In fact, Eq.~(\ref{eq:PP}) can be written in 
a different form 
\begin{eqnarray}
  \label{eq:PP2}
\langle P(x)P(0) \rangle &=& X 
+ Y \left(\frac{1}{V}\sum_{p \neq 0} \frac{e^{ipx}}{p^2+m_\pi^2+\Delta m_\pi^2}\right)
+\cdots ,\nonumber\\
\Delta m_\pi^2 &=& -Z/Y -m_\pi^2,
\end{eqnarray}
of which difference from the original Eq.~(\ref{eq:PP}) 
is the next-to-next-to-next-to-leading order (NNNLO).
By a direct calculation of the zero mode \cite{Suzuki},
one can show
\begin{eqnarray}
\lim_{m\Sigma V \to \infty} \Delta m_\pi^2 &=&0,\;\;\;\;\;
 \lim_{m\Sigma V \to 0} \Delta m_\pi^2 = \frac{2}{F^2 V}.
\end{eqnarray}
This expression suggests that if we can remove 
$X$ and $Y$, the remaining correlator looks almost
the same as that in the conventional $p$ regime,
except for a perturbative correction to the pion mass.
Note that even though the relative correction to the pion mass is ${\cal O}(1)$ 
its influence to the correlator is small since
the conditions $p^2\gg m_\pi^2$ and $p^2\gg \Delta m_\pi^2$ 
are kept for any $p^2$ in a finite volume.

We proceed as follows.
First, we insert a spatial momentum ${\bf p}$ and subtract the 
correlator at a different time-slice $t_{\rm ref}$ in the case of ${\bf p}={\bf 0}$:
\begin{eqnarray}
  C^{\rm 2pt}_{PP}(t,{\bf p})\equiv
  \int d^3x e^{-i{\bf p}\cdot{\bf x}}
  \langle P(x)P(0) \rangle, \\
  \Delta_t C^{\rm 2pt}_{PP}(t,{\bf 0}) \equiv C^{\rm 2pt}_{PP}(t,{\bf 0})-
  C^{\rm 2pt}_{PP}(t_{\rm ref},{\bf 0}),
\end{eqnarray}
and then, take a ratio of them,
\begin{eqnarray}
\frac{C^{\rm 2pt}_{PP}(t,{\bf p})}{\Delta_t C^{\rm 2pt}_{PP}(t,{\bf 0})}
&=& 
\frac{E({\bf 0})\sinh(E({\bf 0})T/2)}{E({\bf p})\sinh(E({\bf p})T/2)}\times
\frac{\cosh(E({\bf p})(t-T/2))}
{\cosh(E({\bf 0})(t-T/2))-
\cosh(E({\bf 0})(t_{\rm ref}-T/2))},
\end{eqnarray}
where $E({\bf p})=\sqrt{{\bf p}^2+m_\pi^2+\Delta m_\pi^2}$.
Note that this ratio is finite even in the 
limit $E({\bf 0})=0$.
We can thus eliminate the leading zero-mode's contributions
$X$ and $Y$ in (\ref{eq:PP2}).
Here, $t_{\rm ref}$ should be taken as large as possible
to avoid the contamination from the excited states,
provided the data at $t_{\rm ref}$ is statistically reliable.


\begin{figure*}[tb]
  \centering
  \includegraphics[width=12cm]{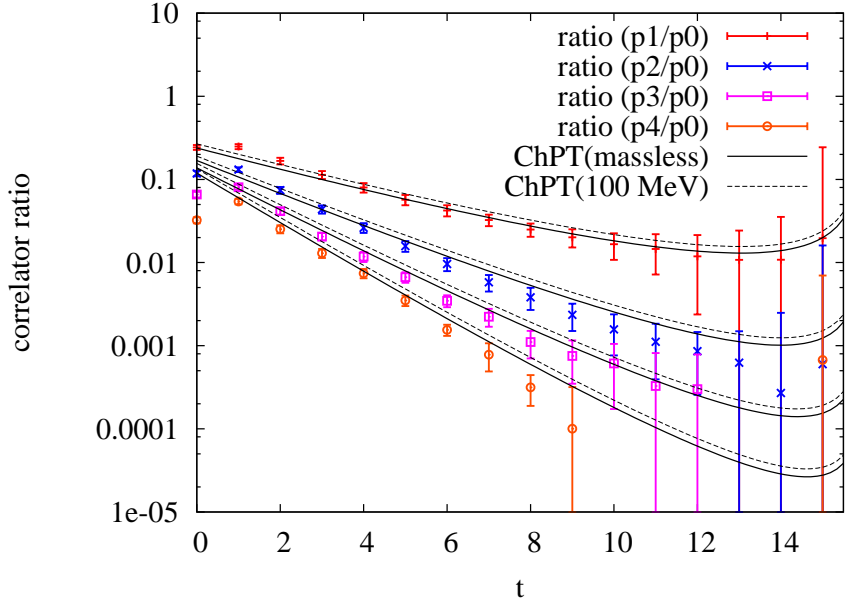}
  \caption{Ratio,
    $C^{\rm 2pt}_{PP}(t,{\bf p})/\Delta_t C^{\rm 2pt}_{PP}(t,{\bf 0})$,
    of two-point correlators at different momenta where we set $t_{\rm ref}=16$.
    Lattice data are plotted together with the expectation of ChPT at
    the leading order without mass (ignoring $Z$) (solid curve) and
    with mass $(m_\pi^2+\Delta m_\pi^2)^{1/2}=100$ MeV (dashed curve).
    Different symbols represent ${\bf p}$ = (1,0,0), (1,1,0), (1,1,1)
    and (2,0,0) in units of $2\pi/L$.
    Here, the rotationally symmetric correlators are averaged in the data.
  }
  \label{fig:2pt}
\end{figure*}

In order to validate the idea based on the form (\ref{eq:PP2}) we
make a plot of the ratio
$C^{\rm 2pt}_{PP}(t,{\bf p})/\Delta_t C^{\rm 2pt}_{PP}(t,{\bf 0})$
with $t_{\rm ref}=16$, for different momenta ${\bf p}$ in Figure~\ref{fig:2pt}, and 
compare with its expectation at the leading order in the
$\epsilon$ expansion in solid (neglecting $Z$) and dashed
(we input $(m_\pi^2+\Delta m_\pi^2)^{1/2}=100$ MeV) curves.
The solid curves, at the leading-order of ChPT neglecting $Z$,
have  no free parameter to tune, as they are simply constructed from massless propagators. 
The agreement of the lattice data with the expectation is fairly good.
Also, we can see that the difference from the massive correlators
is tiny, although some deviations are seen in higher momentum correlators,
which may imply momentum dependent 
higher-order corrections from non-zero modes.
This good agreement indicates that the dominant part of the
finite size effect or the peculiarity of the $\epsilon$ regime
is eliminated, and the remaining non-trivial NLO contribution coming
from the $Z$ term is small compared to the statistical fluctuation.

\section{Three-point functions in the $\epsilon$ regime}
\label{sec:3pt}


In this section, we apply the idea of eliminating the zero-mode
contribution to the three-point function.  For this purpose, we first
consider the $\epsilon$ expansion of the pion form factor within the
framework of ChPT, since the finite volume effect is dominated by the
lightest degrees of freedom, {\it i.e.} the pion.  In the physical
pion form factor, on the other hand, the contribution beyond the
leading terms of the $\epsilon$ expansion becomes important as
suggested by the fact the the vector pole dominance $1/(1-q^2/m_V^2)$
describes the data quite well \cite{Aoki:2009qn, Kaneko:2010ru}. 
Such higher order contributions
are irrelevant to the study of the leading finite volume effect
considered in this work.

It is straight-forward to extend the analysis of the two-point
function described in the previous subsection to the case of
three-point function. It is expressed as a series, of which
each term is a product of the constant due to zero-mode integrals and
the massless propagators of the $\xi$ field such that they connect to
form the three-point function. When the propagator must carry non-zero
momentum, the constant term due to the zero-mode integral cannot
arise.

In the case of our 
``pseudoscalar--(zero-component) vector--pseudoscalar'' correlator,
the series is expressed by
\begin{eqnarray}
  \label{eq:3pt}
  \langle P(x)V_0(y)P(z)\rangle  &=&
  A \;\frac{1}{V}\sum_{p \neq 0} \frac{ip_0}{p^2} 
  \left(e^{ip(x-y)}+ e^{ip(y-z)}\right)
  \nonumber\\&&
  \hspace{-1.3in}
  +B \;\frac{1}{V^2}\sum_{p \neq 0}\;\sum_{p^\prime \neq 0}
  \frac{(ip_0+ip_0^\prime)e^{ip(x-y)}e^{ip^\prime(y-z)}}{p^2p^{\prime 2}}
  F_V((p-p^\prime)^2)
  +\cdots ,
\end{eqnarray}
where $F_V(q^2)$ denotes the vector form factor of the pion
(which is equivalent to our target electro-magnetic form factor when
the up and down quarks are degenerate),  
and $A$, $B$, $\cdots$ are dimensionful constants, including the
contributions from the pion zero-mode.

Inserting an initial (spatial) momentum ${\bf p}_i$ to $P(x)$, 
and a final momentum ${\bf p}_f$ to $P(z)$, we define a three-point
function 
\begin{eqnarray}
  C^{\rm 3pt}_{PVP}(t,t^\prime ; {\bf p}_i, {\bf p}_f) 
  &\equiv&\int d^3x \;e^{-i{\bf p}_i\cdot{\bf x}} \int d^3z \;e^{i{\bf p}_f\cdot{\bf z}}
  \langle P(x)V_0(y)P(z)\rangle,
\end{eqnarray}
where we assume $t=x_0-y_0 < T/2$, and $t^\prime=y_0-z_0 < T/2$.
As in the discussion of the two-point function, we then define a
difference operator 
$\Delta_t f(t)=f(t)-f(t_{\rm ref})$ 
with a fixed value of $t_{\rm ref}$.
It should not be confused with the conventional derivative operator
$\partial_t$. Here, the choice for $t_{\rm ref}$ is more restricted
than in the case of two-point functions, since it should satisfy
both of $t+t_{\rm ref}\ll T$ and $t^\prime+t_{\rm ref}\ll T$
to avoid the contribution of the unusual modes wrapping around the lattice.
In the following analysis, we choose $t_{\rm ref}=T/3=12$,
and use the data at $t<t_{\rm ref}$ and $t^\prime<t_{\rm ref}$.


We construct the following three ratios:
\begin{eqnarray}
  \label{eq:RV1}
  R^1_V(t, t^\prime ; |{\bf p}_i|, |{\bf p}_f|, q^2)
  &\equiv&
  \frac{\displaystyle \frac{1}{N^{\rm 3pt}_{|{\bf p}_i|, |{\bf p}_f|}}
    \sum_{{\rm fixed}|{\bf p}_i|, |{\bf p}_f|, q^2}
    \frac{C^{\rm 3pt}_{PVP}( t, t^\prime ; {\bf p}_i, {\bf p}_f)}
    {E({\bf p}_i)+E({\bf p}_f)}}
  {\displaystyle\left(\frac{1}{N^{\rm 2pt}_{|{\bf p}_i|}}
      \sum_{{\rm fixed}|{\bf p}_i|}C^{\rm 2pt}_{PP}( t,{\bf p}_i)\right)
    \left(\frac{1}{N^{\rm 2pt}_{|{\bf p}_f|}}
      \sum_{{\rm fixed}|{\bf p}_f|}C^{\rm 2pt}_{PP}( t^\prime,{\bf p}_f)\right)},
\end{eqnarray}
\begin{eqnarray}
  \label{eq:RV2}
  R_V^2(t, t^\prime ; |{\bf p}_i|, {\bf 0}, q^2)
  &\equiv&
  \frac{\displaystyle \frac{1}{N^{\rm 3pt}_{|{\bf p}_i|}}
    \sum_{{\rm fixed}|{\bf p}_i|, q^2}
    \Delta_{t^\prime} C^{\rm 3pt}_{PVP}( t, t^\prime ; {\bf p}_i,{\bf 0})
  }
  {\displaystyle\frac{1}{N^{\rm 2pt}_{|{\bf p}_i|}}
    \sum_{{\rm fixed}|{\bf p}_i|}C^{\rm 2pt}_{PP}( t,{\bf p}_i)
    \left[
      -\Delta_{t^\prime} \partial_{t^\prime} C^{{\rm 2pt}}_{PP}( t^\prime ,{\bf 0})
      +E({\bf p}_i)\Delta_{t^\prime} C^{{\rm 2pt}}_{PP}( t^\prime ,{\bf 0})
    \right]},
  \nonumber\\
\end{eqnarray}
\begin{eqnarray}
  \label{eq:RV3}
  R_V^{3}(t, t^\prime ; {\bf 0}, {\bf 0}, q^2=0)
  &\equiv&
  \frac{\displaystyle 
    \Delta_t \Delta_{t^\prime} C^{\rm 3pt}_{PVP}( t, t^\prime ; {\bf 0},{\bf 0})}
  {\displaystyle
    -\Delta_t C^{\rm 2pt}_{PP}(t,{\bf 0})\Delta_{t^\prime} 
    \partial_{t^\prime} C^{{\rm 2pt}}_{PP}( t^\prime ,{\bf 0})
    -\Delta_t \partial_t C^{\rm 2pt}_{PP}(t,{\bf 0})\Delta_{t^\prime}
    C^{{\rm 2pt}}_{PP}( t^\prime ,{\bf 0})},
  \nonumber\\
\end{eqnarray}
where $q^2=({\bf p}_i-{\bf p}_f)^2-(E({\bf p}_i)-E({\bf p}_f))^2$.
Here, the correlators that are equivalent under cubic rotations are
averaged. 
$N^{\rm 3pt}_{|{\bf p}_i|, |{\bf p}_f|}$ and $N^{\rm 2pt}_{|{\bf p}_i|}$ 
denote the numbers of correlators to be averaged.
For $R_V^2$ in (\ref{eq:RV2}) we can interchange the role of initial
and final states and include the case of $|{\bf p}_i|=0$ and 
$|{\bf p}_f|\not=0$.

Using the expression in (\ref{eq:3pt}), it is not difficult to confirm
that these ratios $R_V^{k=1,2,3}$ share the same LO contribution in ChPT, 
{\it i.e.} 
\begin{eqnarray}
  R^{k=1,2,3}_V(t, t^\prime ; |{\bf p}_i|, |{\bf p}_f|, q^2) 
  &=& \frac{B}{Y^2}F_V(q^2)+\cdots.
\end{eqnarray}
Then, one can eliminate the zero-mode contribution $B/Y^2$
by taking their ratios.
Noting $F_V(0)=1$, we can extract the form factor through the ratios
\begin{eqnarray}
  \label{eq:FV1}
  F^1_V(t,t^\prime,q^2)&\equiv& 
  \frac{R^1_V(t, t^\prime ; |{\bf p}_i|, |{\bf p}_f|, q^2)}
  {R_V^{3}(t, t^\prime ; {\bf 0}, {\bf 0}, 0)},
  \\
  \label{eq:FV2}
  F^2_V(t,t^\prime, q^2) &\equiv&
  \frac{R^2_V(t, t^\prime ; |{\bf p}_i|, {\bf 0}, q^2)}
  {R_V^{3}(t, t^\prime ; {\bf 0}, {\bf 0}, 0)}.
\end{eqnarray}
They should become independent of $t$ and $t'$ as long as the ground
state pion dominates the correlator.

So far we have not given an explicit form of  $F_V(q^2)$ 
in the $\epsilon$ expansion since
it may contain the physics at higher orders, as well as those beyond ChPT, as explained above.
In particular, we do not ignore the pion mass, which appears at the higher order in the $\epsilon$ expansion,
in the momentum transfer and simply assume the dispersion relation
of the pion energy: $E({\bf p})=\sqrt{{\bf p}^2+m_\pi^2}$
in the following analysis.
As shown in the previous section, we expect that 
inclusion of the mass should not change the analysis very much, as it is an NLO effect.
The possible distortion of the dispersion relation due to the NLO finite volume effects
will be discussed later.


Because of the zero-mode fluctuation,
there is an unusual contribution which includes 
the scalar form factor of the pion \cite{Fukaya:2014uba}.
But this diagram has a pion propagator
directly connecting the two pseudoscalar sources,
and thus, is expected to be exponentially small.
Since the diagram has different
$t$ and $t^\prime$ dependences of $F^{1,2}_V(t,t^\prime,q^2)$,
we can, in principle, numerically confirm if it is really small or not.
Here, and in the following, we simply ignore this contribution
(we do not observe any unusual $t$ and 
$t^\prime$ dependences of $F^{1,2}_V(t,t^\prime,q^2)$
in the following analysis).

As a final remark of this section, we would like to note
that taking ratios is not a new idea but has been 
widely used for different purposes.
The ratio method non-perturbatively cancels 
the renormalization factors of the operators,
makes the effect of excited modes easier to be detected, 
and so on.
Our work shows the ratio (after inserting momenta)
is also helpful to remove the dominant part of
finite volume effects.

\section{Lattice results}
\label{sec:lattice}

We use gauge configurations of size $16^3\times 48$
generated with the Iwasaki gauge action and 2+1 dynamical flavors of 
overlap quark action.
At $\beta=2.3$, the value of lattice cutoff $1/a$ = 1.759(8)(5)~GeV 
($a\sim$ 0.112(1) fm) is obtained using the $\Omega$-baryon mass as an
input. 
The lattice size in the physical unit is thus $L\sim$ 1.8 fm.

In this work, we focus on an ensemble with the smallest up-down quark
mass, $ma=0.002$, among a set of ensembles with various sea quark
masses. 
This value roughly corresponds to 3~MeV in the physical unit,
and the pion mass at this value is 
$m_\pi \sim$ 99~MeV \cite{Fukaya:2011in},
which is below the physical point. 
For the strange quark, we choose its mass almost 
at the physical value, $m_s a$ = 0.080.
In this set up the pions are in the $\epsilon$ regime
($m_\pi L\sim 0.90$), while kaons remain in the $p$ regime.

Along the Hybrid Monte Carlo simulation, the global topological
charge is fixed at $Q=0$.
Since its effect is encoded in the pion zero-mode,
the $Q$ dependence does not appear in the ratios of our correlators
at the leading order of ChPT.

The correlation functions are calculated using the smeared sources
with the form of exponential function. 
To improve the statistical signal, the so-called all-to-all propagator
technique is employed.
Namely, the low-energy part of the correlator is calculated from 160
eigenmodes of the Dirac operator and averaged over different source
points, while the higher-mode contribution is estimated stochastically 
with the dilution technique \cite{Foley:2005ac}.
For $\Delta_t$ for the zero-momentum correlator, 
we use the reference time-slice at $t_{\rm ref}=12$.
For the derivative operator $\partial_t$, we approximate
it by a simple forward subtraction : $\partial_t f(t)=f(t+1)-f(t)$.

We use 148 configurations sampled from 2500 trajectories of the run.
The auto-correlation time of the correlators 
is different depending on the position and momenta.
The longest one, from the two-point function with zero-momentum,
is around 7 trajectories. 
The statistical errors in the analysis 
are estimated by the jackknife method
after binning data in every 20 trajectories.

\begin{figure*}[tb]
  \centering
  \includegraphics[width=10cm]{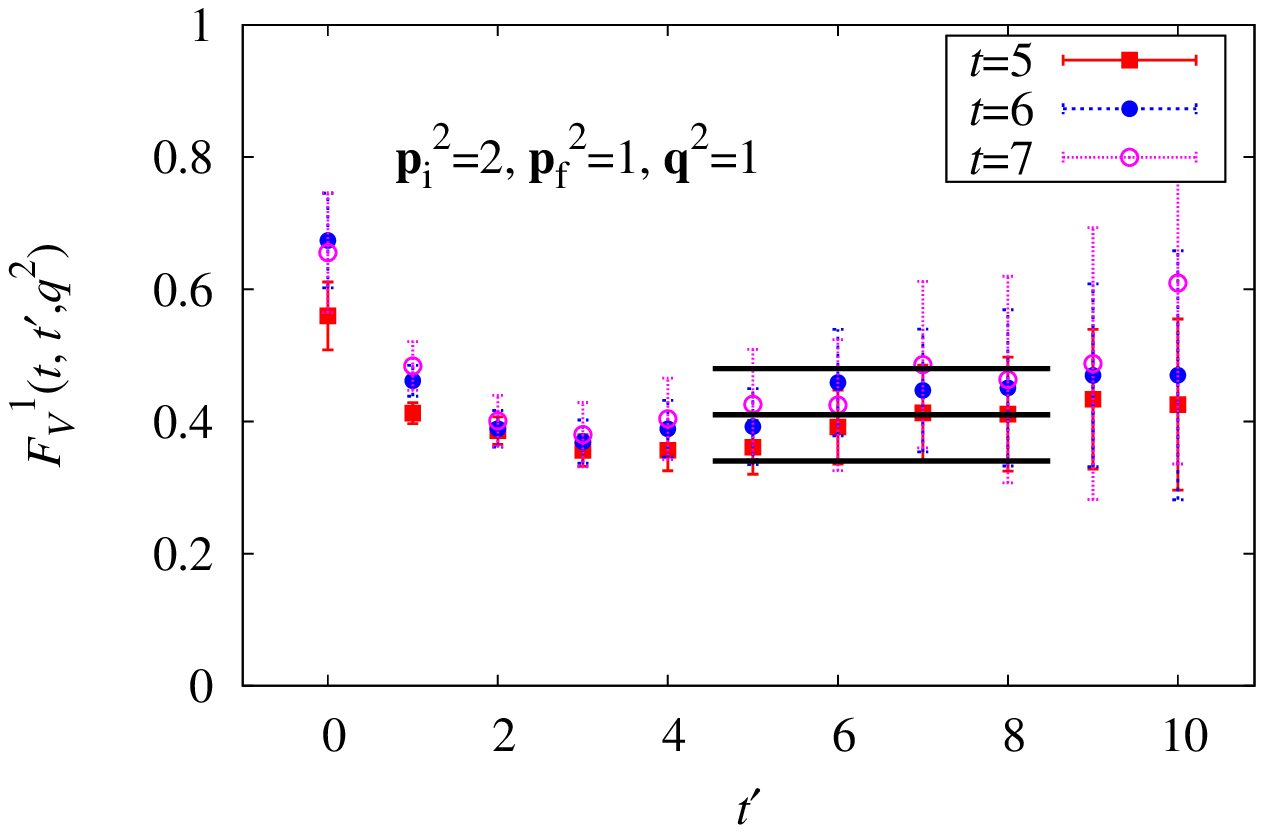}
  \includegraphics[width=10cm]{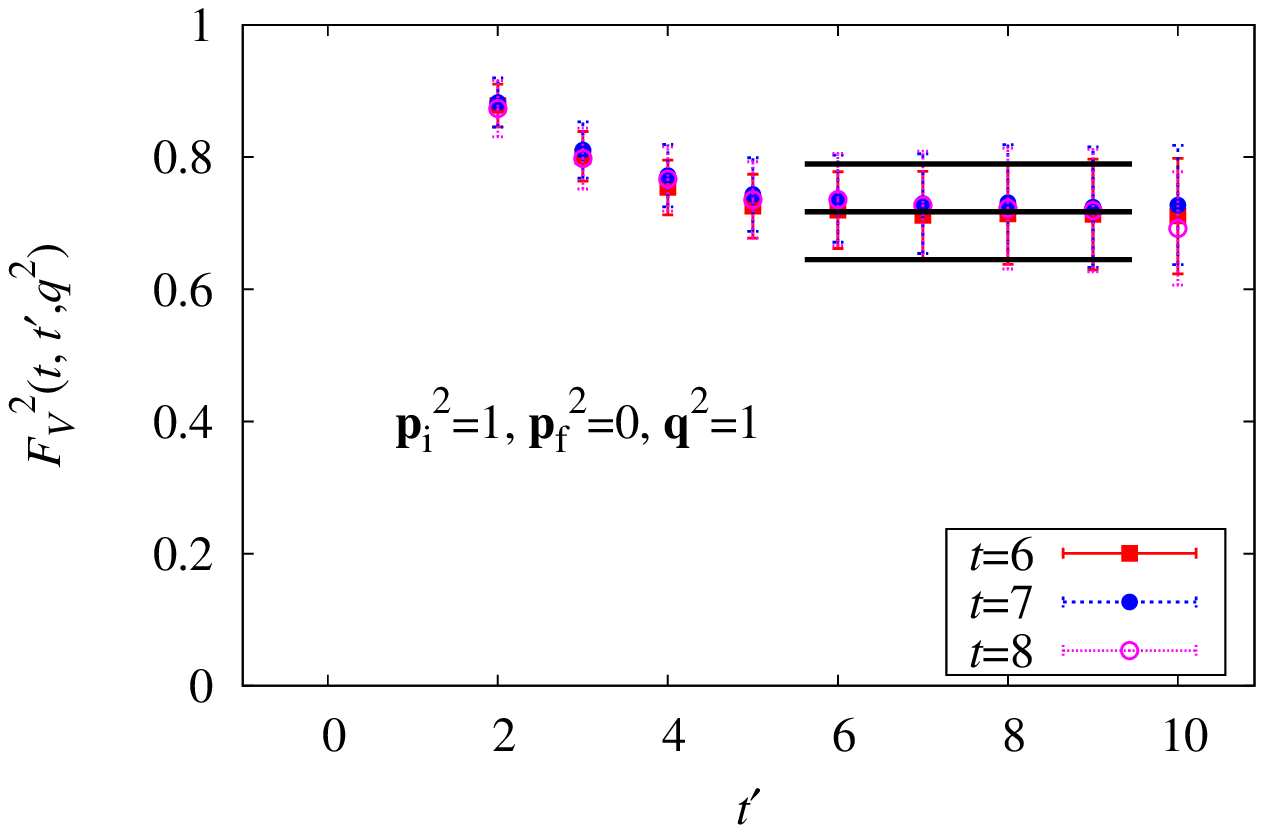}
  \caption{
    $F_V^1(t,t^\prime,q^2=-0.40[\mbox{GeV}^2])$ (top panel) and 
    $F_V^2(t,t^\prime,q^2=-0.11[\mbox{GeV}^2])$ (bottom panel)
    are plotted as a function of $t'$.
    The combination of initial and final momenta are shown in the
    plots in units of $2\pi/L$.
    Data for different $t$'s are plotted with different symbols.
    A constant fit and their error are shown by bands.
  }
  \label{fig:FV}
\end{figure*}

Figure~\ref{fig:FV} presents our lattice data
of $F_V^1(t,t^\prime,q^2)$ at 
$({\bf p}_i^2, {\bf p}_f^2, {\bf q}^2)=(2,1,1)$ 
(top panel) and 
$F_V^2(t,t^\prime,q^2)$ at 
$({\bf p}_i^2, {\bf p}_f^2, {\bf q}^2)=(1,0,1)$
(bottom panel). 
The momenta are labeled in the units of $2\pi /L$.
The ratio $F_V^2$ defined in (\ref{eq:FV2}) is used when either
initial or final spatial momentum is zero with $t_{\rm ref}=12$.
To estimate $E({\bf p})$, which is involved in the definition of
$R_V^1$ in (\ref{eq:RV1}),
we use the dispersion relation $E({\bf p})=\sqrt{{\bf p}^2+m_\pi^2}$.
We find a plateau for time separations where $t$ and $t'$ are greater
than 5, which is also stable against the change of $t_{\rm ref}$ in the range $9\le t_{\rm ref} \le 12$
($t_{\rm ref}$ should satisfy $t+t^\prime+t_{\rm ref}\ll T$).
We fit the data by a constant; the fit results are shown in the plots
as well as the fit range.

\begin{table}
  \centering
  \begin{tabular}{cccc}
    \hline
    $(aq)^2$ & ${\bf p}_i$ & ${\bf p}_f$ & ${\bf q}$\\
    \hline
    0.0380 & (0,0,0) & (1,0,0) & (1,0,0)\\
    0.0560 & (0,0,0) & (1,1,0) & (1,1,0)\\
    0.0699 & (0,0,0) & (1,1,1) & (1,1,1)\\
    0.1281 & (0,1,0) & (1,1,0) & (1,0,0)\\
    0.3084 & (0,$-1$,0) & (1,0,0) & (1,1,0)\\
    0.4366 & (0,0,$-1$) & (1,1,0) & (1,1,1)\\   \hline
  \end{tabular}
  \caption{
    Combinations of initial and final momenta taken in the calculation.
    The momentum components are given in units of $2\pi/L$.
    Those equivalent under cubic rotations are averaged, though not
    listed. 
  }
  \label{tab:q}
\end{table}

\begin{figure*}[tb]
  \centering
  \includegraphics[width=10cm]{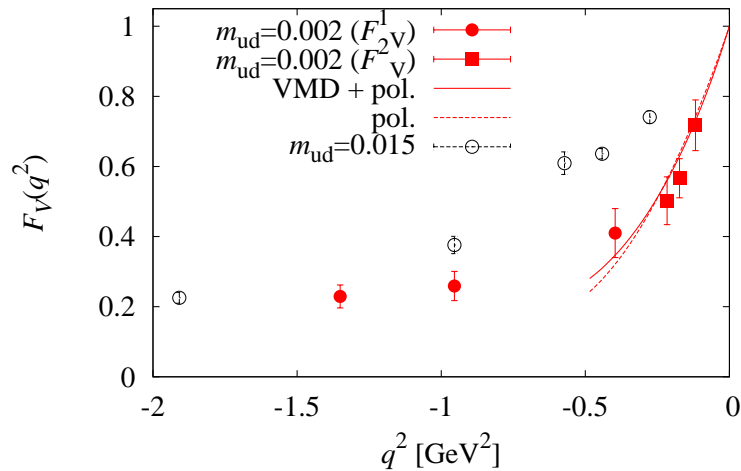}
  \caption{
    Lattice results for $F_V(q^2)$ as a function of $q^2$.
    The lattice data in the $\epsilon$ regime ($m_{ud}a=0.002$, filled
    symbols) and in the $p$ regime ($m_{ud}a=0.015$, open symbols) are
    plotted. 
    The $\epsilon$ regime data are obtained either from the ratio
    $F_V^1$ (circles) or $F_V^2$ (squares).
  }
  \label{fig:FV_vs_q2}
\end{figure*}

We plot $F_V(q^2)$ obtained at various $q^2$ in
Figure~\ref{fig:FV_vs_q2}. 
They are obtained at various combinations of
${\bf p}_i$ and ${\bf p}_f$ listed in Table~\ref{tab:q}.
For comparison, we also plot the data in the $p$ regime 
(at $ma = 0.015$) \cite{Kaneko:2010ru}. 
Apparently, the new data in the $\epsilon$ regime show a steeper slope
near the origin, which indicates a larger value of the pion charge
radius.

We fit the form factor $F_V(q^2)$ to a function
\begin{equation}
  \label{eq:FVfit}
  F_V(q^2)=\frac{1}{1-q^2/m_V^2} + a_1 q^2 +a_2 (q^2)^2,
\end{equation} 
which is motivated by the vector dominance hypothesis and
corrections are added as a polynomial.
The same function was also used in our previous analysis of the
$p$ regime data \cite{Aoki:2009qn}.
Since our calculation of the vector meson mass on the
$\epsilon$ regime ensemble is too noisy to be useful, we use the
physical $\rho$ meson mass, 770~MeV, 
as an input to (\ref{eq:FVfit}) and treat $a_1$ and $a_2$ as
free parameters.
The fit curve goes through the lowest four $|q^2|$ points as shown in
Figure~\ref{fig:FV_vs_q2}.
The $\chi^2/\mbox{dof}$ is below 1.0 in this case.
When we include higher $|q^2|$ points, the fit becomes worse and
$\chi^2/\mbox{dof}$ increases up to 2.5.

The result for the charge radius at our simulated mass is
\begin{equation}
  \label{eq:r2mu002}
  \langle r^2\rangle_V = 0.63(08)(11)\;\;\mbox{fm}^2\;\;\;
  (\mbox{at $m=0.002$}), 
\end{equation}
where the first error is statistical and the second is systematic, as
explained below.
The central value is larger than the experimental value, $0.452(11)$fm$^2$.


Although the main part of the pion zero-mode's effects is removed, 
the systematic error due to finite volume remains the dominant
one. 
First, since the momentum space is discrete, the number of data points
near $q^2=0$ is limited.
The choice of the fitting range and/or fitting function
affects the determination of the slope at $q^2=0$ by 12\%.
Here we assign the variation of the fit results as the systematic
effect. 
(The central value is taken from the fit of the lowest four $|q^2|$
point to (\ref{eq:FVfit}).)
In addition to the model function (\ref{eq:FVfit}), we attempt a
simple polynomial function of second order (dashed curve 
in Figure~\ref{fig:FV_vs_q2}) in this analysis.

Second, the dispersion relation may be distorted in the
$\epsilon$ regime.
By an estimate at the next-to-leading order ChPT, it can be shown that
a distortion of the form
$E({\bf p})\to \sqrt{{\bf p}^2+ Z_m m_\pi^2}$
with $Z_m\sim$ 2 is expected \cite{Fukaya:2014uba}.
Since the relation $E({\bf p})\to \sqrt{{\bf p}^2+  m_\pi^2}$
is used when constructing $R_V^1$, as given in (\ref{eq:RV1}),
this effect may induce a bias as large as $\sim$ 10\%.

Finally, the effect of non-zero modes appeared to be non-negligible
on our small lattice \cite{Fukaya:2014uba}.
We estimate its size as 8\%.
In total, we assign 17\% as the total size of the systematic error by
adding these sources in quadrature.
This is shown in (\ref{eq:r2mu002}) as our estimate of the systematic
error. 
Because it is large, other sources, such as those from discretization
effect, are expected to be subdominant.

\begin{figure*}[tb]
  \centering
  \vspace{0.25in}
  \includegraphics[width=10cm]{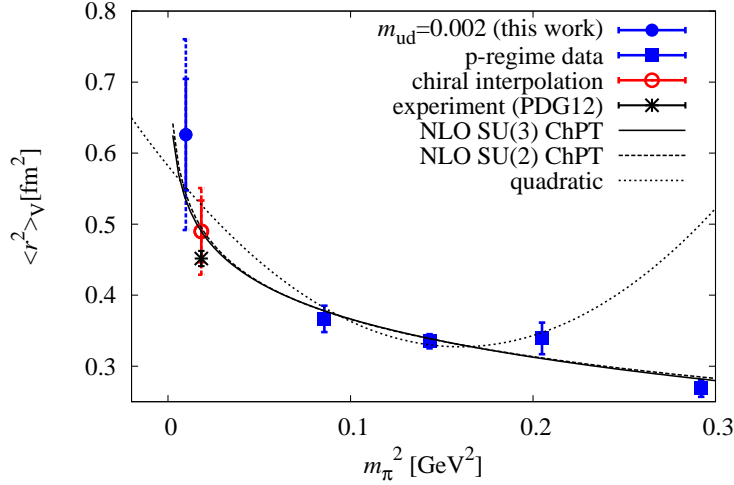}
  \caption{
    Pion charge radius as a function of $m_\pi^2$.
    A result from this work obtained in the $\epsilon$ regime (circle) is
    plotted together with the data at heavier up and down quarks
    (square, from \cite{Kaneko:2010ru}).
    The experimental result is shown at $m_\pi$ = 135~MeV.
    The fit curves are those of next-to-leading order ChPT as well as
    a simple polynomial (quadratic).
  }
  \label{fig:r2_vs_mpi2}
\end{figure*}

Figure~\ref{fig:r2_vs_mpi2} shows the dependence of 
$\langle r^2\rangle_V^\pi$
on the pion mass squared.
The result (\ref{eq:r2mu002}) is plotted together with our previous
calculation at heavier pions \cite{Kaneko:2010ru}.
It is clear that the $\epsilon$ regime result (circle) is much higher
than the points above $m_\pi\gtrsim$ 300~MeV (square).
It indicates the existence of the strong (logarithmic)
curvature of the pion charge radius near the chiral limit.

Finally, we {\it interpolate} the data to the physical pion mass.
We use the functions suggested by the $SU(2)$ and $SU(3)$ ChPT.
At the next-to-leading order, they are
\begin{equation}
  \langle r^2\rangle_V^\pi =
- \frac{1}{NF^2}(1+6N \ell_6^r) - 
  \frac{1}{NF^2}\ln\frac{m_\pi^2}{\mu^2},
\end{equation}
and
\begin{equation}
  \langle r^2\rangle_V^\pi =
  \frac{1}{2NF^2}(-3+24N L_9^r) - 
  \frac{1}{NF^2}\ln\frac{m_\pi^2}{\mu^2} -
  \frac{1}{2NF^2}\ln\frac{m_K^2}{\mu^2},
\end{equation}
for $SU(2)$ and for $SU(3)$, respectively,
with $N=(4\pi)^2$ and $F$ the pion decay constant in the chiral limit.
The parameters $\ell_6^r$ 
and $L_9^r$ are relevant low-energy
constants in $SU(2)$ and $SU(3)$ ChPT, respectively.
The result for the charge radius at the physical pion mass is
\begin{eqnarray}
  \langle r^2\rangle_V &=& 0.49(4)(4)\;\;\mbox{fm}^2\;\;\;(\mbox{at physical point}),
\end{eqnarray}
where the first error is statistical and the second is systematic,
including the one in (\ref{eq:r2mu002})
as well as the variation due to the choice of the chiral fit functions.
That includes the ChPT formulas and a polynomial function at the
second order. 

Through the ChPT fits, we also obtain $F$ and $\ell^r_6$ (or $L_9^r$). 
The value for $F$ is lower than its physical value:
57(8)(10)~MeV and 60(9)(9)~MeV for the $SU(2)$ and $SU(3)$ fits.
This is consistent with our previous extensive analysis of the pion
mass and decay constant in two-flavor QCD \cite{Noaki:2008iy}.
The low-energy constants, in the conventional notations, we obtained in this analysis are
\begin{eqnarray}
\bar{\ell}_6 = -6 N \ell_6^r (\mu=m_\pi)&=& 7.5(1.3)(1.5),\\
L_9^r(\mu=\mbox{770 MeV}) &=& 2.4(0.8)(1.0) \times 10^{-3}.
\end{eqnarray}
These values are also smaller than their phenomenological estimates,
which may indicate that NNLO corrections are not negligible,
in our $p$ regime data points.

\section{Discussions and conclusions}
\label{sec:conclusion}
In this work, we propose a method to calculate the pion form factor in
the $\epsilon$ regime.
Inserting momenta to the operators, and taking appropriate ratios of
them, we can eliminate the dominant contribution from the pion
zero-mode. 
A tree-level analysis of the vector pion form factor
in the $\epsilon$ regime confirms this observation;
the result for the pion charge radius is consistent with the
experiment, showing the existence of a logarithmic divergence towards
the chiral limit. 

This cancellation of the zero-mode occurs only at the leading order,
and there should be non-trivial corrections at the next-to-leading
order. 
This remaining finite volume effects are turned out to be sizable in
the presented calculation.
On the lattice of size $L\sim$ 3~fm or larger, such effect would be
reduced to a few \% level.
One may also use the twisted boundary condition for the valence quarks, 
although we need a study of the partially quenched effect in 
the $\epsilon$ regime analysis of ChPT.

We thank P.~H.~Damgaard and T.~Suzuki for useful discussions.
Numerical simulations are performed on the IBM System Blue Gene
Solution at High Energy Accelerator Research Organization
(KEK) under a support of its Large Scale Simulation Program (No. 10-11, 12-06, 12/13-04).
This work is supported in part by the Grant-in-Aid of the
Japanese Ministry of Education
(No. 21674002, 25287046, 25400284, 25800147, 26247043), 
the Grant-in-Aid for Scientific
Research on Innovative Areas (No. 2004:20105001),
and MEXT SPIRE (Strategic Programs for Innovative Research) Field 5 and 
JICFuS (Joint Institute for Computational Fundamental Science).

\end{document}